\pdfoutput=1
\documentclass[5p]{elsarticle}
\journal{Nucl. Instr. and Meth. A}
\usepackage{amsmath,amssymb,amsfonts} % Typical maths resource packages
\usepackage{hyperref}                 % For creating hyperlinks in cross references
\DeclareMathOperator{\R}{Re}

\begin{document}

\begin{frontmatter}

\title{Measurement of a time-periodic magnetic field by rotating coil}
\author{V.~Marusov}
\ead{V.Marusov@gsi.de}
\address{
  GSI Helmholtzzentrum f\"ur Schwerionenforschung GmbH, Darmstadt, Germany
}

\begin{abstract}
%The rotating coil method is well suited for measurement of a static magnetic field. 
A novel technique of the measurement data processing is developed which allows to apply 
the rotating coil method for measurement of a dynamic magnetic field, periodic in time.
The developed technique allows to obtain \mbox{time-dependent} multipoles in 
a single measurement which takes place during one power cycle of the magnet or the 
coil rotation period, whichever is longer.
\end{abstract}


\end{frontmatter}
%
%_________________________________________________________________________________________
\section{Introduction}
The two-dimensional field harmonic expansion and rotating coil magnetometer are known 
for many years and are described in many books, reviews and accelerator schools. We refer 
to Jain \cite{ref:CAS_Magnetic_98_05_A},\cite{ref:CAS_Magnetic_98_05_B} 
and quote a few results which are needed for description of the proposed technique of the 
measurement data processing.
%_________________________________________________________________________________________
\subsection{Plane circular multipoles}\label{sec:circ_multipoles}
For most practical purposes the magnet field analysis for accelerator applications 
can be reduced to a two-dimensional problem and the field can be represented by the 
power series expansion (see \cite{ref:CAS_Magnetic_98_05_A}, sections 1-2): 
\begin{equation}\label{eqn:B_complex_expansion}
 \pmb{B}(z) = B_{y}+i B_{x} 
            = \sum_{n=1}^{\infty} c_{n} \left(\frac{z}{r_{ref}} \right)^{n-1}
\end{equation}
where $B_{x}$ and $B_{y}$ are the \mbox{$x$-} and \mbox{$y$-components} of the magnetic 
flux density, $z = x+i y$ and $r_{ref}$ is a reference radius. The complex expansion 
coefficients \mbox{$c_{n}=b_{n}+i a_{n}$} in commonly accepted jargon are called 
``harmonics'' or ``multipoles''. 

%_________________________________________________________________________________________
\subsection{Measurement of a static field}\label{sec:stat_field_meas}
The flux $\Phi(\phi)$ through an arbitrary coil array has the form (see \cite{ref:CAS_Magnetic_98_05_B}, eqn.30)
\begin{equation}\label{eqn:Flux_static}
 \begin{split} 
  \Phi(\phi) &= \R\sum_{n=1}^{\infty} \underbrace{s_{n} c_{n}}_{\gamma_n} e^{i n\phi} \\
             &= \frac{1}{2}\sum_{n=1}^{\infty}  
                \gamma_n e^{i n\phi} + \text{{\bf (c.c.)}} 
 \end{split} 
\end{equation}
\eject
In eqn. (\ref{eqn:Flux_static}) $s_n$ are complex ``sensitivity factors'', $\phi$ is 
the angle of rotation, ``\mbox{{\bf (c.c.)}}'' means complex conjugate. We introduced 
variables $\gamma_n = s_{n} c_{n}$ which we refer later as ``reduced multipoles'' 
in order to make further calculations less cumbersome. 

In the measurement, flux as a function of the rotation angle is evaluated through the time 
integral of induced voltage $V(t)$ (see \cite{ref:CAS_Magnetic_98_05_B}, eqns. 7,11), readings 
of the voltage integrator are triggered by an angular encoder. In terms of time as a function 
of the rotation angle $t(\phi)$ 
\begin{equation}\label{eqn:Flux_by_phi_evaluation}
 -\int\limits_{t(0)}^{t(\phi)}V(t) dt = 
  \int\limits_{t(0)}^{t(\phi)}\frac{d\Phi}{dt} dt = 
  \int\limits_{t(0)}^{t(\phi)}d\Phi = \Phi(\phi)-\Phi(0)
\end{equation}
As the $\Phi(\phi)$ is evaluated in the measurement, $\gamma_m$ and consequently 
$c_{m}$ can be found by the Fourier transform:
\begin{equation}\label{eqn:gamma_evaluation}
 \begin{split}
 \frac{1}{\pi}\int\limits_0^{2\pi}\Phi e^{-im\phi}\,d\phi &= 
 \frac{1}{2\pi}\sum_{n=1}^{\infty}
    \gamma_n\underbrace{\int\limits_0^{2\pi} e^{in\phi} e^{-im\phi}\,d\phi}_{2\pi\delta_{nm}}+ \\
    &\quad\quad\quad\; 
    +\overline{\gamma_n}\underbrace{\int\limits_0^{2\pi} e^{-in\phi} e^{-im\phi}\,d\phi}_{0}
 \\
 &= \gamma_m = s_m c_m
\end{split}
\end{equation}
%

%_________________________________________________________________________________________
\section{Measurement of a time-periodic field}\label{sec:meas_periodic}
Let us consider a magnet which is supplied by a periodic current. If the field is 
measured by the rotating coil method, two cases may be separated, which are considered 
in the following sections.

%_________________________________________________________________________________________
\subsection{``Fast coil'' case}\label{sec:meas_periodic_fast_coil}
We assume the measuring coil rotates uniformly making an integer number of turns $M$ 
during the magnet cycling period $T$. 
Let us use $\tfrac{T}{2\pi}$ as the time unit, consequently the dependence of the 
``reduced multipole'' $\gamma_n$ on time can be represented by a Fourier series:
\begin{equation}\label{eqn:multipole_Fourier}
 \gamma_n(t) = \sum_{k=0}^{\infty}\sigma_{nk} e^{ikt}
\end{equation}
and the flux through the coil vs. time is:
\begin{equation}\label{eqn:periodic_flux}
 \Phi(t) = \frac{1}{2}\sum_{n=1}^{\infty}\sum_{k=0}^{\infty}\sigma_{nk} e^{i(k+Mn)t}
         + \text{{\bf (c.c.)}}
\end{equation}
The Fourier transform of (\ref{eqn:periodic_flux}) yields
\begin{equation}\label{eqn:periodic_flux_inv_Fourier}
 \frac{1}{\pi}\int\limits_0^{2\pi}\Phi\,e^{-i\widehat{m}t}\,dt = 
 \sum_{n=1}^{\infty}\sum_{k=0}^{\infty}\delta_{\widehat{m},(k+Mn)}\sigma_{nk}
\end{equation}
Let us consider a partial sum of (\ref{eqn:periodic_flux}) with $n\leq N$, $k < K$ and 
map the $\sigma_{nk}$ to a vector $\mathbf{z}$ of length $NK$ as follows:
\begin{equation}\label{eqn:z_vector_compose}
 \begin{aligned}
   z_{k+(n-1)K} = \sigma_{nk}\quad &\Leftrightarrow\quad 
   z_j = \sigma_{\text{floor}\left(\frac{j}{K} \right)+1,\;j\bmod K} \\
   \quad 0 &\leq j < NK
 \end{aligned}
\end{equation}
In the above equation $\text{floor}(j/K)\equiv (j - (j\bmod K))/K$ is the integer division. 
The equation $k+nM=\widehat{m}$ in terms of the linear index $j$ defined in (\ref{eqn:z_vector_compose})
can be derived as
\begin{equation}\label{eqn:m=k+nM_in_terms_of_j}
 (j \bmod K)(K-M)+jM=(\widehat{m}-M)K
\end{equation}
The smallest solution of the eqn. (\ref{eqn:m=k+nM_in_terms_of_j}) is found for $\widehat{m}=M$, $j=0$.
Let us denote $m=\widehat{m}-M$. Eqn. (\ref{eqn:periodic_flux_inv_Fourier}) can then be rewritten in 
the form $\mathbf{Az}=\mathbf{b}$ where the matrix $\mathbf{A}$ and vector $\mathbf{b}$ are defined as:
\begin{equation}\label{eqn:A_b_compose_complex}
 \begin{aligned}
  A_{mj} &= \delta_{(j\bmod K)(K-M)+jM,\; mK} \\
  b_m\;  &= \frac{1}{\pi}\int\limits_0^{2\pi}\Phi\,e^{-i(m+M)t}\,dt
 \end{aligned}
\end{equation}
The case $K=M$ is a trivial one, because $\mathbf{A}$ turns to the identity matrix, yielding $z_m=b_m$. 

In a case $K>M$, which is of more practical importance, the structure of matrix $\mathbf{A}$ 
is more complicated. Let us consider as a case study $\mathbf{A}$ for $K=3$, $M=1$ and $N=3$:
\begin{equation}\label{eqn:case_study_matrix}
 \mathbf{A}\left(
            \begin{aligned}
	      K &= 3, \\
	      M &= 1, \\
	      N &= 3 
            \end{aligned}
           \right) = 
 \left[
  \begin{array}{ccccccccc}
  \multicolumn{1}{c|}{1} & & & & & & & & \\ \cline{1-2}\cline{4-4}
  & \multicolumn{1}{|c|}{1} & & \multicolumn{1}{|c|}{1} & & & & & \\ \cline{2-5}\cline{7-7}
  & & \multicolumn{1}{|c|}{1} & & \multicolumn{1}{|c|}{1} & & \multicolumn{1}{|c|}{1} & & \\ \cline{3-3}\cline{5-8}
  & & & & & \multicolumn{1}{|c|}{1} & & \multicolumn{1}{|c|}{1} & \\ \cline{6-6}\cline{8-9}
  & & & & & & & & \multicolumn{1}{|c}{1} 
  \end{array}
 \right]
\end{equation}
One can see the matrix contains $KN-(K-M)(N-1)=K+M(N-1)$ rows. It is filled with a regular 
pattern made of $N$ parallel to the diagonal groups of ones, each group of length $K$. 
Positions of ones in $n^{\text{th}}$ group are given by equation $m=j-(n-1)M$. 
Each column in the matrix contains exactly one ``1'', which means that each unknown enters 
only into one equation, thus all equations are decoupled. 

As the number of equations (rows) is less than the number of unknowns, the system must 
be supplemented by more equations to have a unique solution. These additional equations can be 
obtained using the translation properties of $\sigma_{nk}$ and $b_m$. 

Each equation with $L > 1$ terms has, after mapping back from $z_j$ to $\sigma_{nk}$, 
the following form:
\begin{equation}\label{eqn:single_eqn_form}
 \sum_{l=0}^{L-1} z_{j_{\text{min}}+l(K-M)} = b_m
 \;\Leftrightarrow\;
 \sum_{l=0}^{L-1}\sigma_{n_{\text{min}}+l,\;k_{\text{max}}-Ml} = b_m
\end{equation}
In equation (\ref{eqn:single_eqn_form}) $j_{\text{min}}$ is the minimal index with \mbox{$z$-mapping} 
of unknowns (\ref{eqn:z_vector_compose}), $n_{\text{min}}$ and $k_{\text{max}}$ are  
the minimal value of index $n$ and the maximal value of index $k$ in the corresponding 
$\sigma_{nk}$ set, respectively.

If the measurement clock is shifted by $\tau$, from (\ref{eqn:multipole_Fourier}) 
and (\ref{eqn:A_b_compose_complex}) it follows that $\sigma_{nk}$ and $b_m$ are transformed as:
\begin{equation}\label{eqn:sigma_nk_translation}
 \begin{aligned}
  \sigma_{nk} &\;\rightarrow\;\tilde{\sigma}_{nk} = e^{ik\tau}\sigma_{nk} \\
  b_{m}       &\;\rightarrow\;\,\tilde{b}_m \,= e^{i(m+M)\tau} b_{m}
 \end{aligned}
\end{equation}
In terms of the equation for $z_j$, from the original equation (\ref{eqn:single_eqn_form}) we 
can construct another one:
\begin{equation}\label{eqn:z_equation_transform}
 \begin{aligned}
  \sum_{l=0}^{L-1} z_{j_{\text{min}}+l(K-M)} &= b_m \quad\rightarrow \\
  \rightarrow\; \sum_{l=0}^{L-1} e^{-ilM\tau}\cdot z_{j_{\text{min}}+l(K-M)} 
                &= e^{i[m+M-(j_{\text{min}}\bmod K)]\tau} b_m
 \end{aligned}
\end{equation}
Let us construct $L-1$ additional equations as in (\ref{eqn:z_equation_transform}) with 
different $\tau_l$ with $l < L$. Denoting $p_l = e^{-iM\tau_l}$  and setting $p_0=1$ we see 
that the matrix $\mathbf{V}$ built from rows of coefficients of each equation is a square  
Vandermonde matrix \cite{ref:Vandermond_matrix} with a known determinant:
\begin{equation}\label{eqn:Vandermond_matrix}
 \begin{aligned}
 \mathbf{V} &= 
  \begin{bmatrix} 
   1      & p_0     & p_0^2     & \cdots & p_0^{L-1} \\
   1      & p_1     & p_1^2     & \cdots & p_1^{L-1} \\
   1      & p_2     & p_2^2     & \cdots & p_2^{L-1} \\
   \vdots & \vdots  & \vdots    & \ddots & \vdots    \\
   1      & p_{L-1} & p_{L-1}^2 & \cdots & p_{L-1}^{L-1}
  \end{bmatrix} 
 \\
 \left|\mathbf{V}\right| &= \prod_{0\leq i < j < L} (p_j - p_i)
 \end{aligned}
\end{equation}
If all $p_l$ are chosen to be different (a natural choice: $p_l = e^{-2\pi i\frac{l}{L}}$), 
$\mathbf{V}$ has a \mbox{non-zero} determinant, therefore it has an inverse, which for 
the Vandermonde matrix can be explicitly expressed \cite{ref:Vandermond_matrix_inverse}.

Applying the above technique to the case study we obtain the following set of equations:

\begin{equation}\label{eqn:case_study_solution}
\begin{aligned}
  z_0 &= b_0 \\
  \left[
  \begin{array}{rr}
   1 &  1 \\
   1 & -1
  \end{array}
  \right]
  \left[
  \begin{array}{l}
   z_1 \\ z_3
  \end{array}
  \right]
  &=
  \left[
  \begin{array}{r}
   b_1 \\ -b_1
  \end{array}
  \right] \\
  \left[
  \begin{array}{lll}
   1 & 1 & 1 \\
   1 & e^{-i\frac{2\pi}{3}} & e^{-i\frac{4\pi}{3}} \\
   1 & e^{-i\frac{4\pi}{3}} & e^{-i\frac{8\pi}{3}} 
  \end{array}
  \right]
  \left[
  \begin{array}{l}
   z_2 \\ z_4 \\ z_6
  \end{array}
  \right]
  &=
  \left[
  \begin{array}{r}
   b_2 \\ e^{i\frac{2\pi}{3}}b_2 \\ e^{i\frac{4\pi}{3}}b_2
  \end{array}
  \right] \\
  \left[
  \begin{array}{rr}
   1 &  1 \\
   1 & -1
  \end{array}
  \right]
  \left[
  \begin{array}{l}
   z_5 \\ z_7
  \end{array}
  \right]
  &=
  \left[
  \begin{array}{r}
   \;b_3 \\ \;b_3
  \end{array}
  \right] \\
  z_8 &= b_4 \\
 \end{aligned}
\end{equation}
As all matrices in (\ref{eqn:case_study_solution}) are invertible, the case study is solved. 
This technique can be applied for any $N$, $M$ and $K>M$.

Instead of inverting the $NK\times NK$ matrix, the problem is reduced to inverting 
of a number of matrices of smaller dimensions. It can be shown for the general case, that 
the maximum matrix dimension does not exceed $(K-M+1)\times(K-M+1)$.

%_________________________________________________________________________________________
\subsection{``Fast magnet'' case}\label{sec:meas_periodic_fast_magnet}
Let us assume the measuring coil rotates uniformly and the magnet is cycled $M$ times
during one turn of the coil. In this case we denote as $T$ the coil rotating period and measure time 
in units of $\tfrac{T}{2\pi}$. For the purpose of a partial reuse of results obtained in section 
\ref{sec:meas_periodic_fast_coil} let us swap the meaning of indices: now $k$ stands for the 
multipole number and $n$ for term number in the Fourier expansion of the multipole:
\begin{equation}\label{eqn:multipole_Fourier_M}
 \gamma_k(t) = \sum_{n=0}^{\infty}\sigma_{kn} e^{iMnt}
\end{equation}
The flux through the coil vs. time is then given by equation which is similar to 
(\ref{eqn:periodic_flux_inv_Fourier}), except for the low limits of sums:
\begin{equation}\label{eqn:periodic_flux_M}
 \Phi(t) = \frac{1}{2}\sum_{n=0}^{\infty}\sum_{k=1}^{\infty}\sigma_{kn} e^{i(k+Mn)t}
         + \text{{\bf (c.c.)}}
\end{equation}
The Fourier transform of (\ref{eqn:periodic_flux_M}) yields
\begin{equation}\label{eqn:periodic_flux_inv_Fourier_M}
 \frac{1}{\pi}\int\limits_0^{2\pi}\Phi\,e^{-i\widehat{m}t}\,dt = 
 \sum_{n=0}^{\infty}\sum_{k=1}^{\infty}\delta_{\widehat{m},(k+Mn)}\sigma_{kn}
\end{equation}
Let us consider a partial sum of (\ref{eqn:periodic_flux_M}) with $n < N$, $k \leq K$ and 
map the $\sigma_{nk}$ to a vector $\mathbf{z}$ of length $NK$ as follows:
\begin{equation}\label{eqn:z_vector_compose_M}
 \begin{aligned}
   z_{k-1+nK} = \sigma_{kn}\quad &\Leftrightarrow\quad 
   z_j = \sigma_{j\bmod K+1,\;\text{floor}\left(\frac{j}{K} \right)} \\
   \quad 0 &\leq j < NK
 \end{aligned}
\end{equation}
The equation $k+nM=\widehat{m}$ in terms of the linear index $j$ defined in 
(\ref{eqn:z_vector_compose_M}) can be derived as
\begin{equation}\label{eqn:m=k+nM_in_terms_of_j_M}
 (j \bmod K)(K-M)+jM=(\widehat{m}-1)K
\end{equation}
The smallest solution of the eqn. (\ref{eqn:m=k+nM_in_terms_of_j_M}) is found for $\widehat{m}=1$, $j=0$.
Let us denote $m=\widehat{m}-1$, then (\ref{eqn:periodic_flux_inv_Fourier_M}) can be rewritten in 
the form $\mathbf{Az}=\mathbf{b}$ where the matrix $\mathbf{A}$ and vector $\mathbf{b}$ are defined as:
\begin{equation}\label{eqn:A_b_compose_complex_M}
 \begin{aligned}
  A_{mj} &= \delta_{(j\bmod K)(K-M)+jM,\; mK} \\
  b_m\;  &= \frac{1}{\pi}\int\limits_0^{2\pi}\Phi\,e^{-i(m+1)t}\,dt
 \end{aligned}
\end{equation}
Comparing (\ref{eqn:A_b_compose_complex_M}) and (\ref{eqn:A_b_compose_complex}) one can 
see that the same matrix  $\mathbf{A}$ is obtained (albeit after swapping meanings of $k$ and $n$). 
Therefore the consideration of the ``fast magnet'' case is very similar to the one of 
``fast coil''.

The case $K=M$ is a trivial one, because $\mathbf{A}$ turns to the identity matrix, yielding $z_m=b_m$. 

For $K>M$ each equation with $L > 1$ terms has, after mapping from $z_j$ to $\sigma_{kn}$, 
the following form:
\begin{equation}\label{eqn:single_eqn_form_M}
 \sum_{l=0}^{L-1} z_{j_{\text{min}}+l(K-M)} = b_m
 \quad\Leftrightarrow\quad
 \sum_{l=0}^{L-1}\sigma_{k_{\text{max}}-Ml,\;n_{\text{min}}+l} = b_m
\end{equation}
If the measurement clock is shifted by $\tau$, $\sigma_{kn}$ and $b_m$ are transformed as:
\begin{equation}\label{eqn:sigma_nk_translation_M}
 \begin{aligned}
  \sigma_{kn} &\;\rightarrow\;\tilde{\sigma}_{kn} = e^{inM\tau}\sigma_{kn} \\
  b_{m}       &\;\rightarrow\;\,\tilde{b}_m \,= e^{i(m+1)\tau} b_{m}
 \end{aligned}
\end{equation}
In terms of equation for $z_j$, from the original equation (\ref{eqn:single_eqn_form_M}) we 
can construct another one:
\begin{equation}\label{eqn:z_equation_transform_M}
 \begin{aligned}
   \sum_{l=0}^{L-1} z_{j_{\text{min}}+l(K-M)} &= b_m \quad\rightarrow \\
   \rightarrow\; \sum_{l=0}^{L-1} e^{ilM\tau}\cdot z_{j_{\text{min}}+l(K-M)} &= 
     e^{i\left[m+1-M\cdot\text{floor}\left(\frac{j_{\text{min}}}{K}\right)\right]\tau}\cdot b_m
 \end{aligned}
\end{equation}
Consequently, exactly as for the ``fast coil'' case, from each equation with $L>1$ terms we 
construct $L-1$ additional equations using the $\sigma_{kn}$ and $b_m$ time translation 
properties, obtaining a linear system with a Vandermonde matrix, which has a \mbox{non-zero} 
determinant.

Just as in the ``fast coil'' case, the problem is reduced to inverting a number of matrices with 
dimensions not exceeding $(K-M+1)\times(K-M+1)$.

%_________________________________________________________________________________________
\section{Discussion}\label{sec:discussion}
For the ``fast magnet'' case (Section \ref{sec:meas_periodic_fast_magnet}) a good 
synchronization of the coil rotation and the magnet cycling is essential. As a possible 
implementation one can consider the magnet power supply driven by a waveform generator 
which is clocked by the angular encoder pulses.

Much looser requirements can be applied for the ``fast coil'' case 
(Section \ref{sec:meas_periodic_fast_coil}). 
The magnet cycle starts from the injection plateau and comes back to it. If multipoles 
stay constant at the injection plateau then expulsion of a part of the plateau, followed 
by ``gluing'' of remaining parts preserves continuity of each multipoles as a function of time.

Therefore in the ``fast coil'' case the measurement can be started at any point shortly before 
the ramp up, and ended at any point during the injection plateau after ramp down when the coil 
comes back to the start angle. In other words, the magnet cycle is not required to 
be a multiple of the coil rotation period. This option, to choose a ``fictitious'' magnet 
cycle, may be exploited. It may be shown for a cycle with the same ramp up and ramp down rates,
that the closer the ``fictitious'' injection plateau duration is to the duration of the cycle \mbox{flat-top} 
the lesser are higher harmonics in the Fourier expansion (\ref{eqn:multipole_Fourier}).

The developed data processing technique operates with truncated expansions (partial sums) 
to represent the field and multipoles. To check if upper limits of sums are large enough 
to provide the precision required they may be varied and the effect of variations on obtained 
results must be analyzed.
%\vfill
\eject
%_________________________________________________________________________________________
\section{Conclusions}\label{sec:conclusions}
The data processing technique has been developed which allows the use of the rotating coil method for 
the measurement of a dynamic \mbox{time-periodic} field. 
This technique allows to obtain \mbox{time-dependent} multipoles in 
a single measurement which lasts one power cycle of the magnet or the coil rotation period, 
whichever is longer.

%_________________________________________________________________________________________
\section*{Acknowledgments}\label{sec:acknowledgments}
The author would like to thank Markus Kirk for valuable comments and help with preparation 
of the manuscript.

%%%%%%%%%%%%%%%%%%%%%%%%%%%%%%%%%%%%%%%%%%%%%%%%%%%%%%%%%%%%%%%%%%%%%%%%%%%%%%%%%%%%%%%%%%

\end{document}